# Trajectory Interpretation of Correspondence Principle: Solution of Nodal Issue

Ciann-Dong Yang[1] and Shiang-Yi Han[2]

[1]Department of Aeronautics and Astronautics, National Cheng Kung University, Tainan, Taiwan. R.O.C.

[2]Department of Applied Physics, National University of Kaohsiung, Kaohsiung, Taiwan. R.O.C

**Abstract**

The correspondence principle states that the quantum system will approach the classical system in high quantum numbers. Indeed, the average of the quantum probability density distribution reflects a classical-like distribution. However, the probability of finding a particle at the node of the wave function is zero. This condition is recognized as the nodal issue. In this paper, we propose a solution for this issue by means of complex quantum random trajectories, which are obtained by solving the stochastic differential equation derived from the optimal guidance law. It turns out that point set A, which is formed by the intersections of complex random trajectories with the real axis, can represent the quantum mechanical compatible distribution of the quantum harmonic oscillator system. Meanwhile, the projections of complex quantum random trajectories on the real axis form point set B that gives a spatial distribution without the appearance of nodes, and approaches the classical compatible distribution in high quantum numbers. Furthermore, the statistical distribution of point set B is verified by the solution of the Fokker-Planck equation.

**Keywords: Correspondence principle, Complex quantum random trajectory, Complex Fokker-Planck equation**

[2] Corresponding author, Assistant Professor, Email: syhan.taiwan@gmail.com

# 1. Introduction

The correspondence principle not only plays an important role in quantum mechanics but also holds a very crucial key to connecting the microscopic and macroscopic worlds. It states that quantum mechanics will reduce to classical mechanics within the limits of the quantum number approaching infinity, and has been widely discussed even in the infancy of the quantum era [1-5]. The quantum probability density is in good agreement with the classical probability, on average, with the principal discrepancy being the rapid oscillations in $|\Psi|^2$. These oscillations generate the same number of nodes as the quantum number in Fig. 1 shows. This nodal issue remains unsolved despite the fact that many semiclassical or classical-like interpretations of quantum mechanics have been proposed. A proper interpretation of the correspondence principle is needed as technology starts to transcend the limits of both quantum and classical boundaries. The nodal issue and classical compatible probability distribution are two main challenges encountered by the new interpretation of the correspondence principle.

The correspondence principle has been discussed from different perspectives by different interpretations [6-10]. The agreement between the classical and quantum probability density of the quantum harmonic oscillator improves rapidly with an increase in the quantum number [11]. However, the behavior of the quantum harmonic oscillator in the stationary state differs from the classical prediction [12]. It was pointed out that the correspondence principle cannot be applied to all periodic systems via a demonstration of a particle in a cubical box [13].

Bohmian mechanics (BM) proposes a trajectory interpretation of quantum mechanics on the basis of the pilot wave guidance law [14]. In this trajectory interpretation, the quantum potential is responsible for quantum phenomena [15-17]. Holland showed that classical behavior could take place for some quantum systems when the quantum potential is negligible [18]. Other discussions of the correspondence principle in different approaches refer to [19, 20]. However, a vanishing quantum potential is not a general condition to have at the classical limit [21]. It seems that the Bohmian classical limit can only be realized by combining narrow wave packets, mixed states, and environmental decoherence, as suggested by Bowman [22].

In recent years, the complex quantum trajectory (CQT) interpretation has been applied to some quantum phenomena, such as the tunneling effect [23-25], quantum chaos [26-28], and scattering [29, 30]. The significant benefits brought out by the CQT interpretation are the manifestation of the matter wave [31] and the properties of the interferences [32-34]. The double-slit experiments were investigated as intelligible manifestations for the CQT interpretation either in numerical or experimental manner.

Gondran [35] presented the double-slit interference with an ultra-cold atom, and Sanz [36] utilized polarized light to demonstrate the interference pattern through photon trajectories. The first real-time interference image of a single molecule was captured to approve the wave property of the particle [37]. The first quantitative verification of the equivalence between the trajectory-based statistics and the wavefunction-based statistics on slit-experiments was proposed by an ensemble of the QCTs solved from the same Hamilton equation [38].

When it comes to the complex-valued wavefunction, the definition of the probability density should be modified. Poirier [39] discussed the modified form of the complex-valued probability density function and pointed out that it does not satisfy the continuity equation. John [40-42] showed that the complex-valued probability density is inversely proportional to the square of the particle's complex velocity and some related properties. Bender [43] discussed the mathematical characteristic of the complex-valued probability density. The complex-valued probability density of the resonant tunneling system and harmonic oscillator refer to [44, 45].

As is well-known, the probability distribution in BM has to be initially satisfied with the Schrödinger equation such that the subsequent probability distribution is able to represent the quantum probability distribution [46]. However, the initial probability distribution cannot be known in advance and is even more difficult to be obtained in the complex configuration space. We overcome this troublesome issue by introducing random motion combined with the CQT in the complex plane. This complex quantum random trajectory (CQRT) has been proposed underlying the framework of an optimal guidance law in complex space [47], where a remarkable quantum compatible result was given by an ensemble of CQRTs of the Gaussian wave packet oscillating in harmonic potential. In this study, we apply the CQRT interpretation to the quantum harmonic oscillator in different eigenstates, including high quantum number states. We find out that an ensemble of CQRTs in the complex plane can solve the nodal issue and even approach the classical-like probability distribution in the high quantum number. On the other hand, the point set formed by the intersections of CQRTs with the real axis reproduces the quantum probability distribution. We further solve the corresponding Fokker-Planck equation and compare the solution to the CQRT statistical results. It turns out that the numerical probability density solved from the Fokker-Planck equation is in line with the CQRT statistical result.

This paper is organized as follows. We briefly review the CQT interpretation and its two major issues, including the definition of the complex probability density function and the uncertain initial distribution. The CQRT interpretation will be introduced by the formulation of the quantum guidance law and CQRTs of the Gaussian wave packet will be demonstrated in Section 3. We then demonstrate in Section 4 that the statistical spatial distribution given by an ensemble of CQRTs of the quantum harmonic oscillator can represent the classical and quantum distributions, and solve the nodal issue. Furthermore, we show that the probability distribution solved from the Fokker-Planck equation is consistent with the statistical spatial distribution of the ensemble of CQRTs. Section 5 covers the discussions and conclusions.

## 2. Complex Probability Density

Consider a wave function defined in the complex plane, $\Psi(t,z)$, $z = x + iy \in \mathbb{C}$, $x, y \in \mathbb{R}$. The time evolution of $\Psi(t,z)$ is determined by the complex-valued Schrödinger equation, which is equivalent to the quantum Hamilton-Jacobi equation

$$\frac{\partial S}{\partial t} + \frac{(\nabla S)^2}{2m} + U - \frac{i\hbar}{2m}\nabla^2 S = 0, \qquad (2.1)$$

where $S = S_R + iS_I$ is the complex action function related to the wave function $\Psi(t,z)$ by

$$S(t,z) = -i\hbar \ln \Psi(t,z). \qquad (2.2)$$

According to Hamilton's principle, the momentum of the particle is given by $p = \nabla S$. Thus, from Eq. (2.2), we have

$$\dot{z} = (-i\hbar/m)\nabla \ln \Psi. \qquad (2.3)$$

One can express the real part of the velocity as

$$\dot{x} = \frac{\hbar}{2mi}\frac{\Psi\nabla\Psi^* - \Psi^*\nabla\Psi}{\Psi^*\Psi}. \qquad (2.4)$$

The numerator of Eq. (2.4) is in the form of the probability current that one is familiar with in conventional quantum mechanics, but is complex-valued. The denominator is the complex probability density according to Born's rule,

$$\rho(t,x,y) = \Psi^*(t,x,y)\Psi(t,x,y) = |\Psi(t,x,y)|^2 \qquad (2.5)$$

Please note that the LHS of Eq. (2.4) represents the velocity field on the basis of the quantum Hamilton-Jacobi formalism, and the RHS is recognized as the $\Psi$-field presented by quantum mechanics. This implies that one can connect the trajectory-based probability to the wavefunction-based probability through an ensemble of trajectories. Trajectory-based statistics are required to satisfy the continuity equation,

$$\frac{\partial \rho(t,x,y)}{\partial t} + \nabla \cdot \left(\dot{x}(t,x,y)\rho(t,x,y)\right) = 0, \qquad (2.6)$$

which in principle, can be derived from the imaginary part of the quantum Hamilton-Jacobi equation (2.1),

$$\frac{\partial S_I}{\partial t} + \frac{1}{m}\nabla S_R \nabla S_I - \frac{\hbar}{2m}\nabla^2 S_R = 0, \qquad (2.7)$$

via the relation

$$\rho(t,x,y) = e^{-2S_I/\hbar}. \tag{2.8}$$

It is of interest to see that the continuity equation (2.6) is related to the real part of the complex velocity; however, it is determined by both real and imaginary coordinate information. In addition, the relation (2.8) shows that $\rho(t,x,y)$ is determined by the imaginary part of the complex action function, $S_I$. However, $\rho(t,x,y)$ does not always converge, for example $|\Psi(t,x,y)|^2 \to \infty$ as $y \to \pm\infty$. The indeterminate imaginary factor $y$ disqualifies $\rho(t,x,y)$ from being a probability density function.

The initial probability distribution is the other issue in BM. It was pointed out that unless the assumption $\rho_B(0,x_B^0) = |\Psi_B(0,x_B^0)|^2$ can be deduced from other perspectives, BM is essentially not an ordinary statistical mechanics of a deterministic theory [46]. In the CQT interpretation, more initial conditions can be given by a set of starting points, $(x_0, y_0) = (x, y_j)|_{j=1,2,3,\cdots}$, which have a fixed real part $x$ and varying imaginary part $y_j$ of the initial positions. One can have many different initial positions in the complex configuration space. However, no matter how the starting point set is given, it always has to satisfy the initial probability distribution

$$\rho(0, x_0, y_0) = |\Psi(0, x_0, y_0)|^2 \tag{2.9}$$

such that the continuity equation (2.6) can be satisfied. Requirement (2.9) is a fundamental postulate that cannot be proved within the CQT interpretation. In addition, it is very difficult to set the imaginary part of the initial position in that the actual initial position distribution cannot be obtained. A theory that can propose a formalism connecting classical statistics to the quantum probability is fundamentally needed, and it should be equipped with the ability to deal with the issue of the uncertain initial position distribution. Up to now, the initial condition issue remains unsolved. In the following section we will show how to solve this issue by considering random quantum motion.

## 3. Complex Quantum Random Trajectory

From the viewpoint of the control theory, nature itself plays an optimal controller that allows an object to move along the optimal path with minimum energy consumption. In physics, the variational method presents the least action for a deterministic system. For example, when a particle is moving from point A to a fixed point B, its optimal path is determined by the least action. However, if the terminal point is varying due to some randomness, it becomes a stochastic system. To deal with such a stochastic system, we need to apply the dynamic programming method for the reason that only current time and current state are needed. In the complex quantum random trajectory (CQRT) interpretation, it can be seen that the quantum motion of a particle is guided by its accompanying pilot wave in an optimal manner such that the guided motion stems from the minimization of the Lagrangian cost function in the presence of the stochastic Wiener process [47].

Let us consider a stochastic system in the complex plane,
$$dz = u(t,z)dt + \sqrt{v}dw, \tag{3.1}$$
where $v$ represents the diffusion coefficient, $u(t,z)$ is a function to be determined and $w$ is the normalized Wiener process satisfying $E(w) = 0$ and $E(dw^2) = dt$. There are two displacements: $u(t,z)dt$ is the drift displacement, and $\sqrt{v}dw$ represents the random diffusion displacement. To find the optimal function $u(t,z)$, we need to minimize the cost-to-go function:
$$V(t,z) = \min_{u[t,t_f]} J(t,z,u) = E_{t,z}\left\{\int_t^{t_f} L(\tau, z(\tau), u(\tau)) d\tau\right\}, \tag{3.2}$$
where $E_{t,z}\{\cdot\}$ denotes the expectation over all stochastic trajectories starting from $z(t) = z$. An expectation is needed for dealing with the randomness of the cost-to-go function due to the action of noise. Eq. (3.2) can be recast into the stochastic Hamilton-Jacobi Bellman equation:
$$-\frac{\partial V(t,z)}{\partial t} = \min_u\left\{L(t,z,u) + \nabla V(t,z)u(t,z) + \frac{v}{2}\nabla^2 V(t,z)\right\}, \tag{3.3}$$
whose solution is the optimal cost-to-go function, $V(t,z)$. Under the demand of minimizing the terms inside the brace at the fixed time $t$ and the position $z$ when solving Eq. (3.3), the condition
$$\frac{\partial L(t,z,u)}{\partial u} = -\nabla V(t,z), \tag{3.4}$$
determines the optimal command $u^*(t,z)$. Therefore, we have the following stochastic Hamilton-Jacobi-Bellman (HJB) equation
$$-\frac{\partial V(t,z)}{\partial t} = L(t,z,u^*) + \nabla V(t,z)u^*(t,z) + \frac{v}{2}\nabla^2 V(t,z). \tag{3.5}$$
One can derive the Schrödinger equation from the above stochastic HJB equation by choosing $L(t,z,u) = mu^2/2 - U(z)$ as the Lagrangian of a particle with mass $m$ moving in the potential $U(z)$, and $v = -i\hbar/m$ as the diffusion coefficient. For the given Lagrangian $L(t,z,u)$, the optimal drift velocity $u^*$ can be determined from Eq. (3.4) as
$$u^* = -\frac{1}{m}\nabla V(t,z). \tag{3.6}$$
The optimal cost-to-go function $V(t,z)$ is solved from Eq. (3.5) with the above $u^*$:
$$\frac{\partial V}{\partial t} - \frac{1}{2m}(\nabla V)^2 - U - \frac{i\hbar}{2m}\nabla^2 V(t,z) = 0. \tag{3.7}$$
A comparison between Eq. (2.1) and Eq. (3.7) reveals the relation of $V$ to the wave function $\Psi$ as
$$V(t,z) = -S(t,z) = i\hbar \nabla \ln \Psi(t,z). \tag{3.8}$$
It is worthy to notice that the optimal command $u^*$ represents the mean velocity of the random motion described by Eq. (3.1), and is related to the wave function as

$$u^*(t,z) = -\frac{1}{m}\nabla V(t,z) = \frac{1}{m}\nabla S(t,z) = -\frac{i\hbar}{m}\nabla \ln \Psi(t,z). \tag{3.9}$$

While other quantum-trajectory formulations regard $p = \nabla S$ as a fundamental assumption [14-16], it is actually a natural outcome of the optimal guidance law in the CQRT interpretation.

The quantum dynamics of a particle moving randomly in the complex plane is described by the stochastic differential equation (3.1) with the optimal guidance command $u^*$ and the diffusion coefficient $\nu = -i\hbar/m$:

$$dz = u^*(t,z)dt + \sqrt{\nu}dw = -\frac{i\hbar}{m}\nabla\big(\ln\Psi(t,z)\big)dt + \sqrt{-\frac{i\hbar}{m}}dw. \tag{3.10}$$

Eq. (3.10) defines a complex-valued Langevin equation, whose drift velocity $u^*(t,z)$ is in general a nonlinear function of z. We can numerically integrate Eq. (3.10) by using the Euler-Maruyama method with a fixed time step $\Delta t$:

$$z_{j+1} = z_j - \frac{i\hbar}{m}\frac{d\ln\Psi(t_j,z_j)}{dz}\Delta t + \sqrt{\frac{i\hbar}{m}}(1+i)\xi\sqrt{\Delta t},\; j=0,1,\cdots n, \tag{3.11}$$

where $\sqrt{\Delta t}$ stems from the standard deviation of the Wiener process $dw$, and $\xi$ is a real-valued random variable with standard normal distribution $N(0,1)$, i.e., $E(\xi) = 0$ and $\sigma_\xi = 1$. Please note that the CQRT becomes the CQT, if we calculate the expectation value of both sides of Eq. (3.11).

Consider a Gaussian wave packet moving in the complex plane (in dimensionless form),

$$\Psi(t,z) = \frac{1}{\sqrt[4]{\pi}\sqrt{1+it}}\exp\left[ip_0 z - i\frac{p_0^2}{2}t\right]\exp\left[-\frac{(z-p_0 t)^2}{2(1+t^2)}\right]. \tag{3.12}$$

The finite-difference equation (3.11) associated with this wave packet reads

$$z_{j+1} = z_j - i\left(\frac{z_j - p_0 t}{1+t^2}\right)dt + \frac{-1+i}{\sqrt{2}}\xi\sqrt{dt},\; j=0,1,\cdots n, \tag{3.13}$$

which is obtained by inserting the wave function from Eq. (3.12) into Eq. (3.11). All particles are launched from the same initial position $(x_0, y_0) = (0,0)$, and we choose $p_0 = 1$. Fig. 2 demonstrates how the number of trajectories could influence the statistical results. One particle generates one specific trajectory due to the random diffusion displacement even it is launched from the same initial point. The spatial statistical distribution cannot be fulfilled until the number of trajectories is large enough. The spatial distribution of 100,000 trajectories is in good agreement with the quantum probability distribution with the correlation coefficient $\Gamma = 0.9994$, which is shown in Fig. 2d. However, too many trajectories will slightly reduce the correlation coefficient due to the numerical accumulated error. Therefore, we choose 100,000 trajectories to see how the correlation coefficient could increase with time as Fig. 3 illustrates. This shows that the particles need time to spread out to fill the space to reproduce the quantum probability distribution.

In the optimal guidance formalism, we can reproduce the quantum probability distribution by a few initial positions or even a single initial position. In principle, the random motion property allows us to abandon the requirement of the initial probability distribution given by Eq. (2.9). The independence from the initial probability distribution suggests that the CQRT interpretation may in many cases provide a more complete and fundamental formulation with regard to quantum mechanics than the CQT interpretation. In the next section, we will tackle the nodal issue in the quantum harmonic oscillator and interpret the correspondence principle underlying the CQRT interpretation.

## 4. Statistical Distribution of Complex Random Trajectories

The random behavior of the quantum harmonic oscillator is described by Eq. (3.10) in the dimensionless form as

$$dz = -i\frac{\partial \ln \Psi_n(t,z)}{\partial z} dt + \frac{-1+i}{\sqrt{2}} \xi \sqrt{dt}, \quad (4.1)$$

where $\Psi_n(t,z)$ denotes the wave function of the $n^{th}$-state harmonic oscillator. To numerically integrate Eq. (4.1), we rewrite it in the following finite-difference form:

$$z_{j+1} = z_j - i\frac{\partial \ln \Psi_n(t_j, z_j)}{\partial z} \Delta t + \frac{-1+i}{\sqrt{2}} \xi \sqrt{\Delta t}, \quad j = 0, 1, \cdots n. \quad (4.2)$$

Two coupled difference equations arise by separating Eq. (4.2) into the real and imaginary parts:

$$x_{j+1} = x_j + \text{Im}\left(\nabla\left(\ln \Psi_n(t_j, x_j, y_j)\right)\right) \Delta t - \frac{\xi}{\sqrt{2}} \sqrt{\Delta t}, \quad j = 0, 1, \cdots n, \quad (4.3a)$$

$$y_{j+1} = y_j - \text{Re}\left(\nabla\left(\ln \Psi_n(t_j, x_j, y_j)\right)\right) \Delta t + \frac{\xi}{\sqrt{2}} \sqrt{\Delta t}, \quad j = 0, 1, \cdots n. \quad (4.3b)$$

Starting from an arbitrary initial position, Eq. (4.3) can provide an infinite number of CQRTs in the complex plane. For later statistical usage, we introduce two types of point collection sets: point set A and point set B. The former is composed of the intersections of CQRTs with the $x-$axis; while the latter is formed by points which are the projections of CQRTs on the $x-$axis. Fig. 4 gives an illustration of the two point sets.

An ensemble of CQRTs solved from Eq. (4.3) in the $n=1$ state is presented in Fig. 5, which is generated by launching 100,000 particles individually at initial positions $(x_0, y_0) = (\pm 0.95, 0)$ with time step $\Delta t = 0.01$. It is clear to see that CQRTs will spread out with time even with only two initial positions. We obtain point set A by collecting the intersections of CQRTs and the $x-$axis, whose spatial distribution is shown in Fig. 6a. A remarkable consistency with the quantum probability is observed in this figure. The correlation coefficient between the quantum probability and the statistical spatial distribution of point set A is up to $\Gamma = 0.995$. The statistical distribution of the point set A in the $n = 2, 3,$ and $4$ states are demonstrated in Fig. 6b to Fig. 6d, which show strong correlation between the probability-based statistics and the trajectory-based statistics. This observation is intelligible and intuitive since the experimental equipment can only capture the information in the real (actual) space, as collected by point set A.

Now let us consider point set B from the same ensemble of CQRTs in the $n = 1$ state, and compare its spatial distribution with the quantum probability distribution, as shown in Fig. 7a. It is clear to see that the spatial distribution of point set B rules out the nodes, $x = \pm 1$. The nodes of the wave function are the locations where the particle cannot be found in the quantum probability interpretation due to the presence of an infinite quantum potential at the real axis ($x-$axis). For a particle moving along the real axis, the node is a point that cannot be crossed. However, for a particle moving in the complex plane, it can avoid encountering the infinite potential barrier by choosing the other paths with nonzero imaginary components in the complex plane, as shown in Fig. 8. In other words, nodes are no longer present in the distribution of CQRTs. This phenomenon also causes discrepancies between the maximum and minimum of the two distributions in Fig. 7, and gradually transforms the spatial distribution of point set A to the classical probability distribution when the quantum number becomes larger.

We expect the nodes to disappear in the distribution of CQRTs for other eigenstates. Fig. 9 illustrates the spatial distributions of point set B in the states of $n = 10, 30, 50,$ and $70$. What attracts our attention is that the CQRT interpretation not only solves the nodal issue but also recovers the classical probability distribution. From Fig. 9, one notes that at the higher quantum state the particle stays, the better classical-like spatial distribution it generates. Accordingly, the CQRT interpretation is capable of representing a quantum system, which in a statistical sense behaves classically in the high quantum states just like its classical counterpart. The other quantum-classical approach to the quantum harmonic oscillator system given by the CQT interpretation [48] shows that when $n \to \infty$, the quantum potential becomes irrelevant compared to the harmonic potential, and generates an ineffective quantum force, hence, the particle moves classically on the real axis with the imaginary part of motion vanishing. That is to say, the classical-like spatial distribution we established in this paper provides a statistical foundation of the quantum-classical approach given by the CQT interpretation, since CQT is the mean of CQRT.

The probability density function of a stochastic system can be found by solving its accompanying Fokker-Planck equation. The Fokker-Planck equation of the quantum harmonic oscillator with random motions described by the stochastic differential equation (4.1) in the $n = 1$ state reads (in the dimensionless form),

$$\frac{\partial \rho}{\partial t} = \frac{-4xy\rho}{x^2+y^2} + \frac{x^2y+y^3+y}{x^2+y^2}\frac{\partial \rho}{\partial x} + \frac{x-xy^2-x^3}{x^2+y^2}\frac{\partial \rho}{\partial y} + \frac{1}{4}\left(\frac{\partial^2 \rho}{\partial x^2} - \frac{2\partial^2 \rho}{\partial x \partial y} + \frac{\partial^2 \rho}{\partial y^2}\right). \quad (4.4)$$

We solve Eq. (4.4) by using the finite-difference method with the initial condition and the boundary conditions given by

$$\rho(0, x, y) = \frac{2}{\sqrt{\pi}}(x^2 + y^2)e^{-x^2-y^2}, \quad (4.5)$$

$$\rho(t, -L, y) = \rho(t, L, y) = 0, \quad \rho(t, x, -L) = \rho(t, x, L) = 0, \quad (4.6)$$

where $L = 5$ ensures the probability equal to zero on the boundary. Fig. 10a shows that the probability density $\rho$ solved from Eq. (4.4) fits point set B very well with a correlation coefficient $\Gamma = 0.9975$. Fig. 10b illustrates the consistency between the two distributions with $\Gamma = 0.9964$ in the $n = 3$ state. Thus, we have verified the correctness of the CQRTs' spatial distribution by showing its equivalence with the probability distribution solved from the accompanying Fokker-Planck equation.

## 5. Conclusions and Discussions

Quantum mechanics provides the most precise description for the microscopic world. It must be understood that wavefunction does not represent an objective reality, but merely acts as a mathematical tool carrying a particle's information. Quantum mechanics demands that we abandon the classical understanding of reality; however, we must learn from nature itself presented in front of us. In this paper, we propose the CQRT interpretation, which may shed some light on the relationship between the empirical description of nature and its reality revealed in a deterministic sense. The complex configuration space not only acts as a mathematical tool but also retains the ontological interpretation in quantum mechanics. It is essential for a quantum system to transit to a classical system in a high quantum number.

In this study, we have successfully solved the nodal issue in terms of the CQRT interpretation. In addition, we have demonstrated how trajectory-based statistics can approach quantum statistics and classical statistics by adopting different point sets extracted from CQRTs. The reconstruction of the quantum statistics is achieved by means of a collection of intersections of a huge number of CQRTs with the $x$-axis. On the other hand, the projection of an ensemble of CQRTs onto the $x$-axis reproduces the classical distribution in high quantum states. It is emphasized that the convergence of the CQRT distribution to quantum statistics and classical statistics is independent of an initial CQRT distribution. Unlike the existing trajectory interpretations of quantum mechanics, the coincidence of the initial spatial distribution with the quantum probability is not a necessary condition for the CQRT interpretation. This is an advantage of the CQRT interpretation as compared to the BM or the CQT interpretations.

We also verify the probability distribution of CQRTs by showing its equivalence with the distribution solved from the accompanying Fokker-Planck equation. This means that the quantum motion is essentially random and described by stochastic differential equations. In conclusion, we propose the CQRT interpretation to solve the nodal issue of the correspondence principle and to bridge the gap between quantum mechanics and classical mechanics. More quantum systems have to be studied to further verify the effectiveness of the CQRT interpretation.

## Acknowledgments

We thank Tsung-Lien Ko and Yang-Hsuan Lin for performing the numerical simulations.

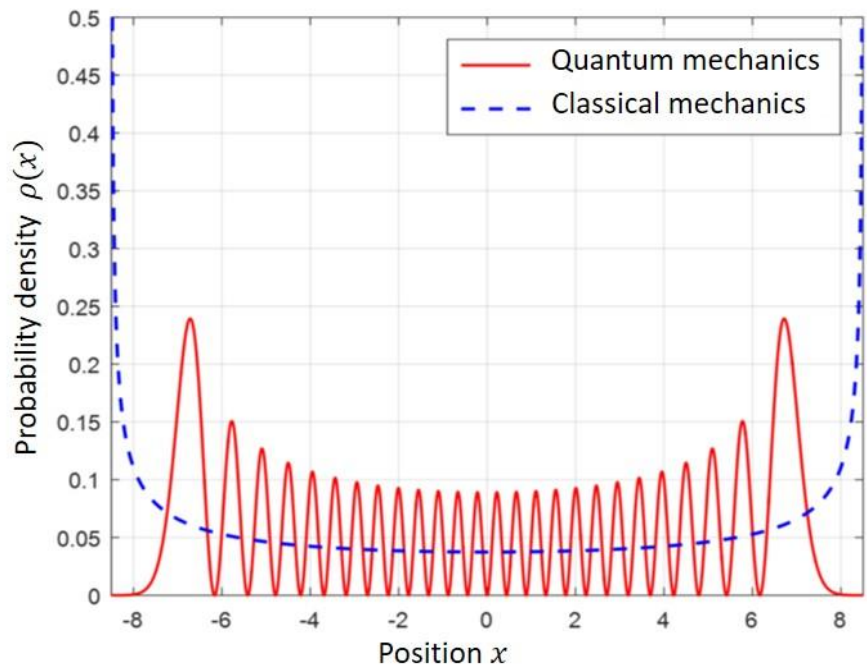

Fig. 1 The comparison of the quantum probability density for the state $n = 25$ of a harmonic oscillator with the classical probability density of the same total energy. A similar plot can be found from Fig.11 in [11].

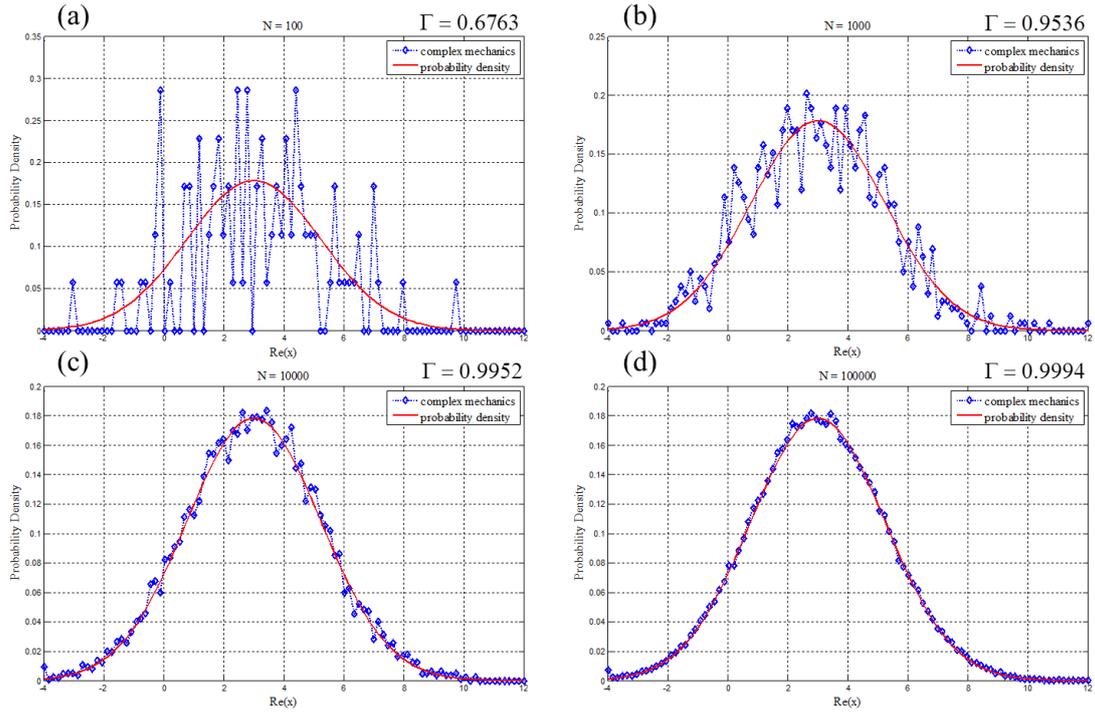

Fig. 2 The comparisons of the quantum probability density (red solid curves) with the spatial distributions generated by different numbers of CQRTs of Gaussian wave packet (blue square points). The number of CQRTs and the resulting correlation coefficients between the two distributions shown in the four subgraphs are given by (a) $N = 100, \Gamma = 0.6763$; (b) $N = 1000, \Gamma = 0.9536$; (c) $N = 10000, \Gamma = 0.9952$; (d) $N = 100,000, \Gamma = 0.9994$.

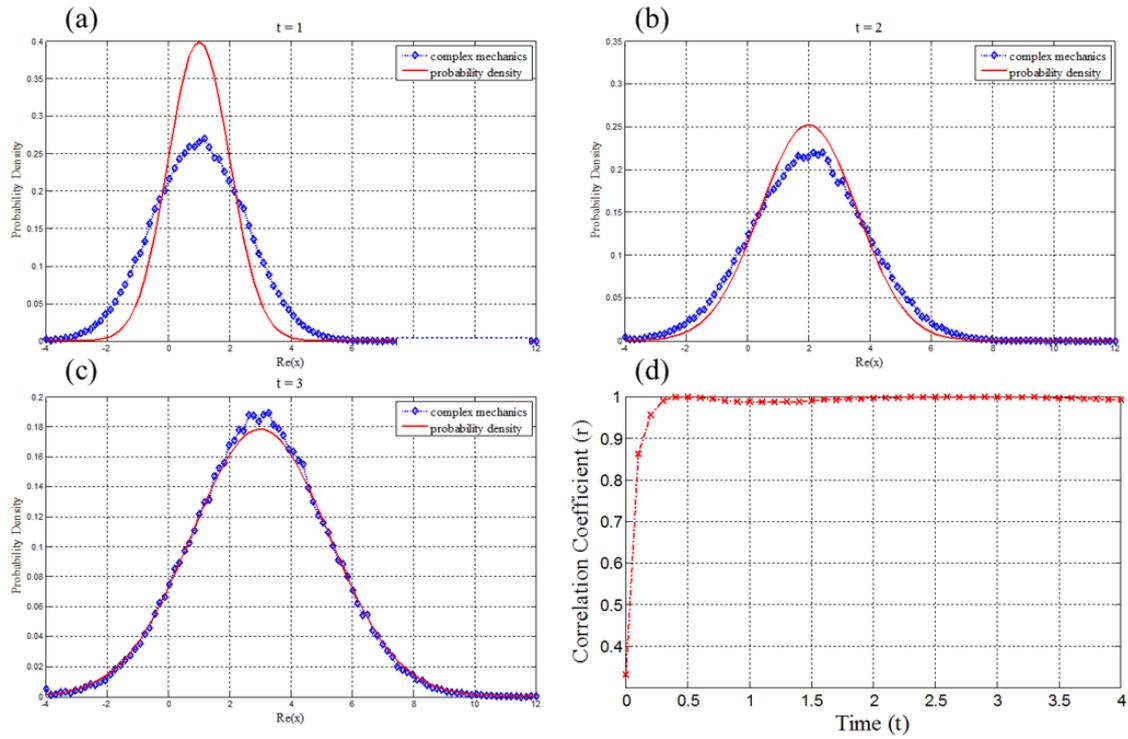

Fig. 3 The comparison of the quantum probability density (red solid curve) with the statistical spatial distribution generated by 100,000 CQRTs (blue square points) of the Gaussian wave packet at three instances $t = 1$, $t = 2$, and $t = 3$. The subgraph (d) illustrates the time evolution of the correlation coefficient between the statistical spatial distribution and the quantum probability.

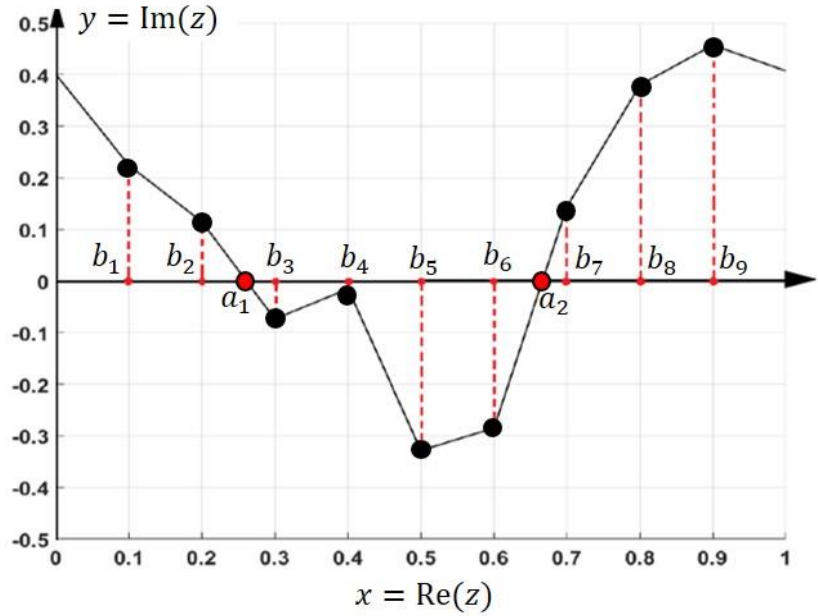

Fig. 4  The two point sets extracted from CQRTs. Point set $A = \{a_1, a_2\}$ is composed of the intersections of CQRTs with the $x-$ axis. Point set $B = \{b_1, b_2, a_1, b_3, b_4, b_5, b_6, a_2, b_7, b_8, b_9\}$ is composed of the projections of CQRTs on the $x-$axis.

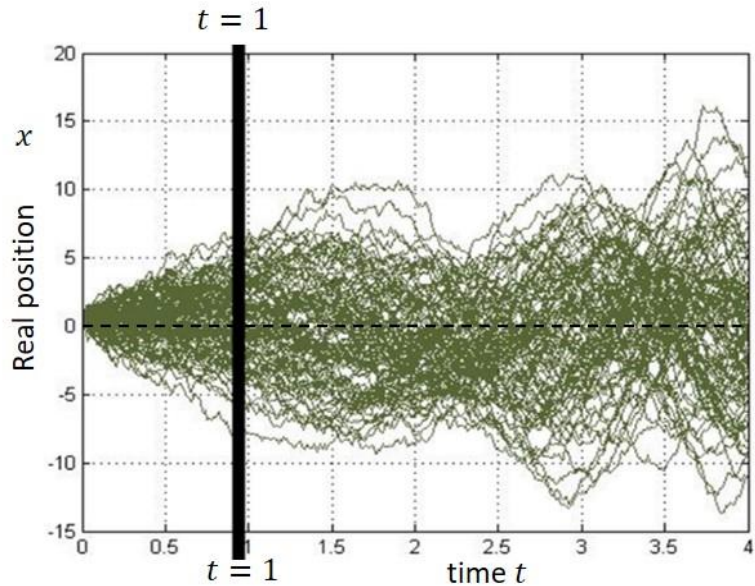

Fig. 5 The time evolution of the real part of 100,000 CQRTs of the quantum harmonic oscillator in the $n = 1$ state. All trajectories start from initial positions $(x_0, y_0) = (\pm 0.95, 0)$.

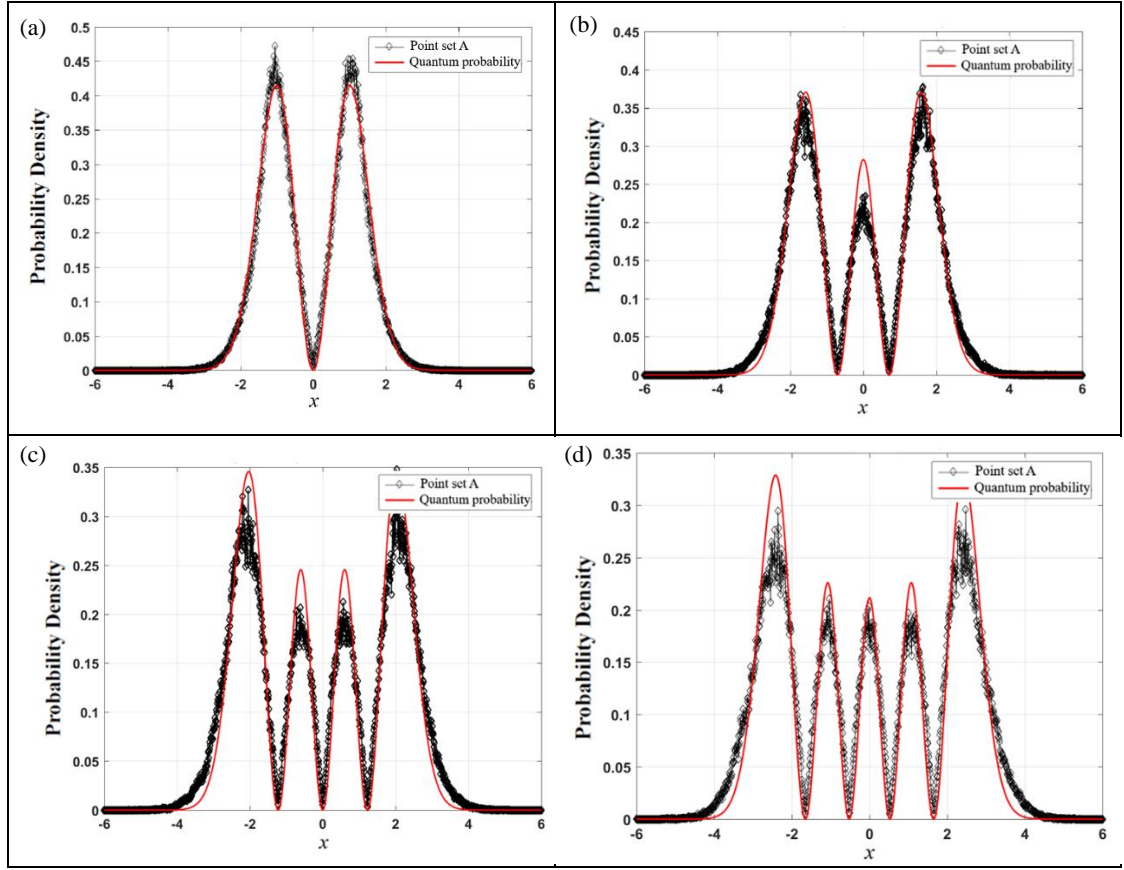

Fig. 6 The comparison of the quantum probability density of the harmonic oscillator (red solid curve) with the statistical spatial distribution of point set A (black circle points) generated by the intersections of 100,000 CQRTs with the $x$−axis at the instance $t = 1$ as shown in Fig. 5. (a) CQRTs in the $n = 1$ state with initial positions $x_0 = \pm 0.95$ yield a correlation coefficient $\Gamma = 0.995$. (b) CQRTs in the $n = 2$ state, with initial positions $x_0 = \pm 1.45, 0$ yield a correlation coefficient $\Gamma = 0.9896$. (c) CQRTs in the $n = 3$ state with initial positions $x_0 = \pm 0.58, \pm 1.88$ yield a correlation coefficient $\Gamma = 0.9808$. (d) CQRTs in the $n = 4$ state with initial positions $x_0 = \pm 1, \pm 2, 0$ yield a correlation coefficient $\Gamma = 0.9880$.

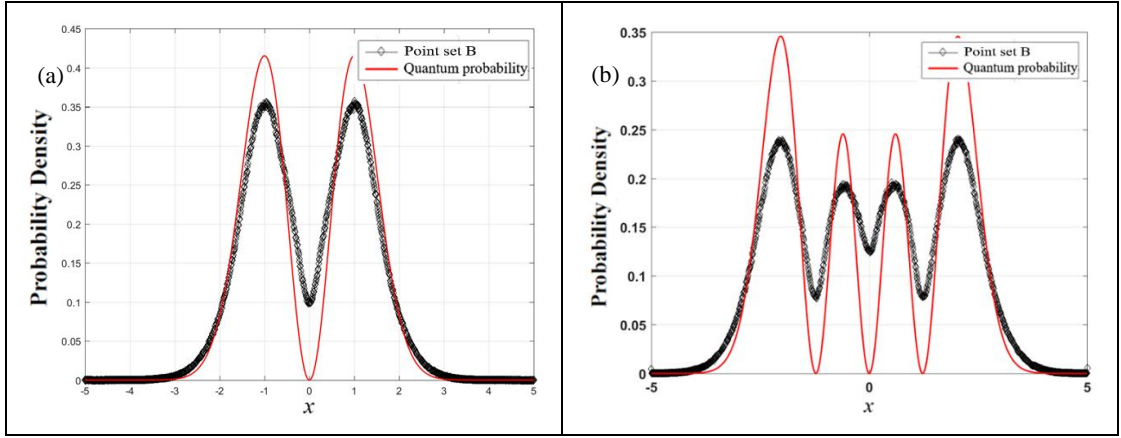

Fig. 7 The comparison of the quantum probability density of the harmonic oscillator (red solid curve) with the statistical spatial distribution of point set B (black circle points) generated by the projection of 100,000 CQRTs onto the $x-$axis. (a) CQRTs in the $n = 1$ state with initial positions $x_0 = \pm 0.95$. (b) CQRTs in the $n = 4$ state with initial positions $x_0 = \pm 1, \pm 2, 0$.

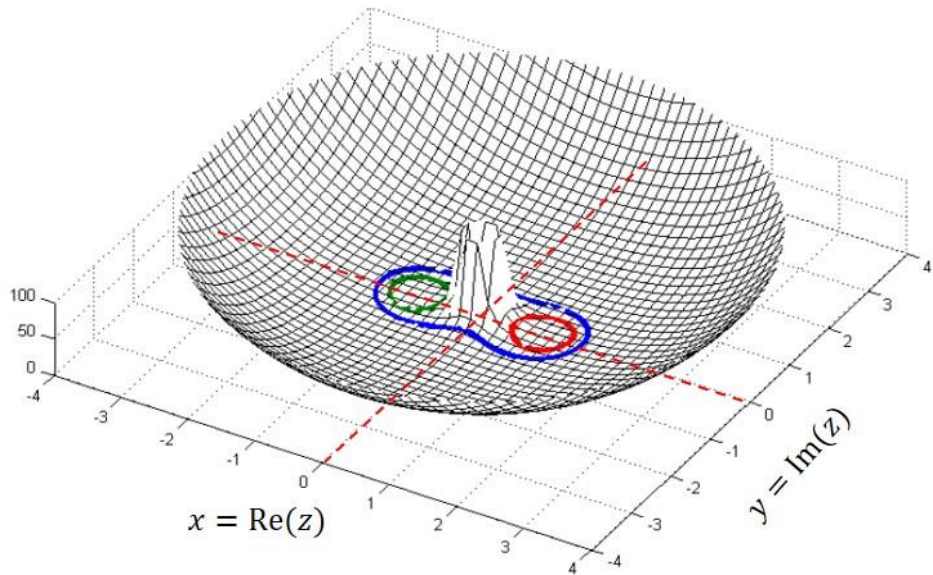

Fig. 8 A particle moving in the complex plane can bypass the node at the origin by moving away from the real axis. Three mean trajectories of the quantum harmonic oscillator in the $n = 1$ state are displayed over the surface of the total potential (including the quantum potential and the harmonic potential). Considering the three trajectories in the figure, point set A excludes the origin, because no trajectory passes the origin. Point set B records the origin two times, because there are two points located at the vertical positions of the blue trajectory, having their projections on the origin.

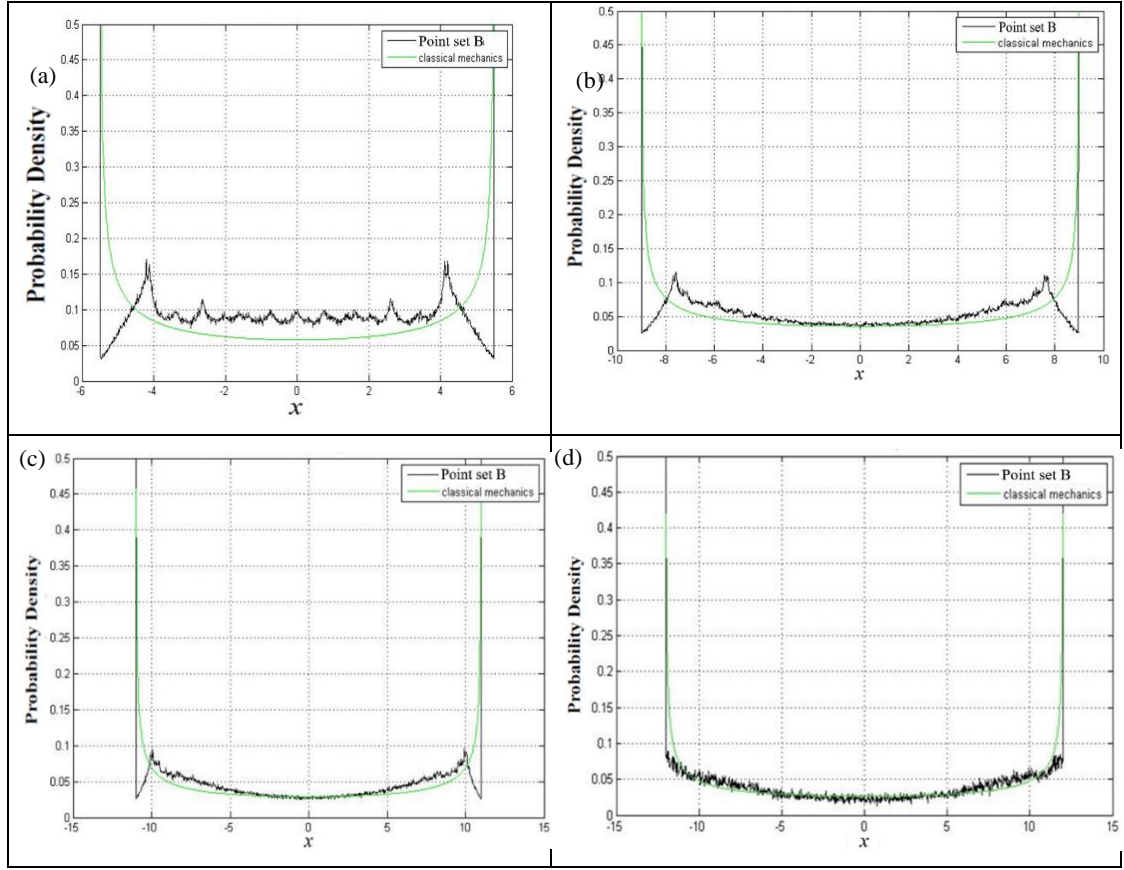

Fig. 9 The comparison of the classical distribution (green curve) with the statistical spatial distributions of point set B (black curve) generated by 100,000 CQRTs projected onto the $x-$axis in different eigenstates of the quantum harmonic oscillator. (a) CQRTs in the $n = 10$ state yield a correlation coefficient $\Gamma = 0.1887$. (b) CQRTs in the $n = 30$ state yield a correlation coefficient $\Gamma = 0.8346$. (c) CQRTs in the $n = 50$ state yield a correlation coefficient $\Gamma = 0.8606$. (d) CQRTs in the $n = 70$ state yield a correlation coefficient $\Gamma = 0.9412$.

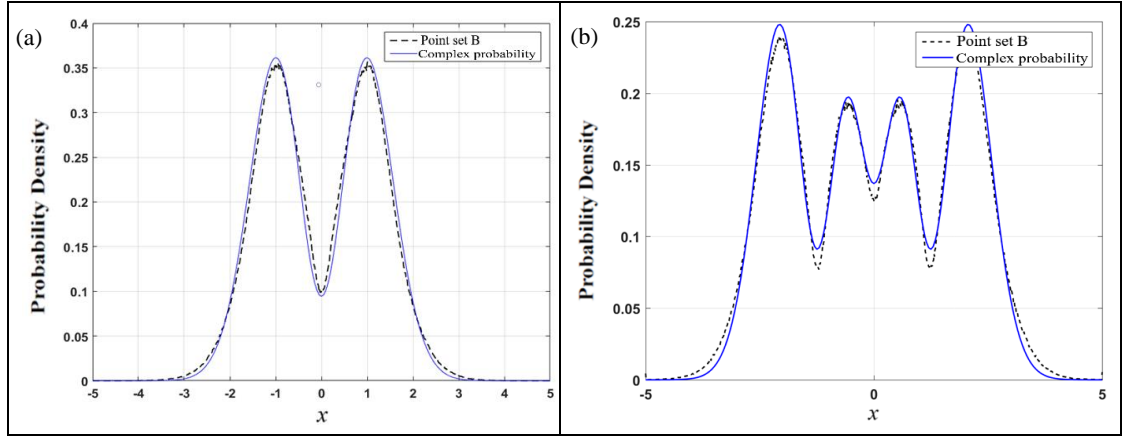

Fig. 10 The probability densities (solid blue curve) solved from the Fokker-Planck equation (4.4) are compared with the statistical spatial distributions given by point set B (dotted black curve). (a) CQRTs in the $n = 1$ state yield a correlation coefficient $\Gamma = 0.9975$. (b) CQRTs in the $n = 3$ state yield a correlation coefficient $\Gamma = 0.9964$.